\documentclass{aa}
\usepackage{graphicx}
\usepackage[varg]{txfonts}
\usepackage{cases}
\usepackage{natbib}

\newcommand{\fei}{Fe \scriptsize{I} \normalsize}

\begin{document}

\title{Observational signature of continuously operating drivers of decayless kink oscillation}

\author{Dong~Li\inst{1,2}, Zhentong~Li\inst{1}, Fanpeng~Shi\inst{1,3}, Yang~Su\inst{1}, Wei~Chen\inst{1}, Fu~Yu\inst{1,3}, Chuan~Li\inst{4,5}, Ye~Qiu\inst{4,5}, Yu~Huang\inst{1}, and Zongjun~Ning\inst{1}}

\institute{Key Laboratory of Dark Matter and Space Astronomy, Purple Mountain Observatory, CAS, Nanjing 210023, China \email{lidong@pmo.ac.cn \& yang.su@pmo.ac.cn} \\
           \and State Key Laboratory of Space Weather, Chinese Academy of Sciences, Beijing 100190, PR China \\
           \and School of Astronomy and Space Science, University of Science and Technology of China, Hefei, 230026, PR China \\
           \and School of Astronomy and Space Science, Nanjing University, Nanjing 210023, China \\
           \and Key Laboratory of Modern Astronomy and Astrophysics (Nanjing University), Ministry of Education, Nanjing 210023, China}

\date{Received; accepted}

\titlerunning{Observational signature of continuously operating drivers of decayless kink oscillation}
\authorrunning{Dong Li et al.}

\abstract {Decayless kink oscillations, which are nearly omnipresent
in the solar corona, are believed to be driven by continuously
operating energy supply.} {In this letter, we investigate an
external continuous excitation of an apparent decayless oscillation
during an X1.1 flare on June 20, 2023 (SOL2023-06-20T16:42).} {The
decayless kink oscillation was identified in the coronal loop at
extreme ultraviolet (EUV) wavelengths and the associated flare
quasi-periodic pulsations (QPPs) were simultaneously observed in
passbands of hard X-ray (HXR), microwave, and ultraviolet (UV)
emissions.} {The kink oscillation is detected as a transverse
oscillation of the coronal loop, which reveals five apparent cycles
with an average period of about 130$\pm$10~s. The oscillation
amplitude does not show any significantly decay, suggesting a
decayless oscillation. At the same time, the solar flare occurs in
the vicinity of the oscillating loop and exhibits five main pulses
in HXR, microwave, and UV emissions, which could be regarded as
flare QPPs. They have similar periods of about 100-130~s, which may
indicate successive and repetitive energy releases during the flare
impulsive phase. The peak of each loop oscillation cycle appears to
follow the pulse of the QPPs, suggesting that the transverse
oscillation is closely associated with flare QPPs.} {Our
observations support the scenario where the repetitive energy
released following flare QPPs could be invoked as external,
continuously operating drivers of the apparent decayless kink
oscillation.}


\maketitle

\section{Introduction}
Kink-mode oscillations are usually identified as transverse
oscillations of loop-like structures and they are always
characterized by non-axisymmetric and weakly compressive in the
long-wavelength regime \citep[see][for a recent
review]{Nakariakov21}. Kink oscillations are well studied
magnetohydrodynamic (MHD) waves, since they play a crucial role in
diagnosing the solar magnetic field and measuring plasma parameters,
namely, "MHD coronal seismology"
\citep[e.g.,][]{Yuan16,Yang20,Chelpanov22}. They are generally
manifested as one of two forms: decaying and decayless oscillations,
which are usually dependent on their oscillation amplitudes. The
decaying oscillation always reveals a large displacement amplitude,
namely, at $\gg$1~Mm, it decays fast and only persists for several
oscillatory cycles.
\citep{Nakariakov99,Su18,Kumar22,Li23a,Li23b,Zhang23}. For the
large-amplitude oscillation, the decay time is about 1.79 times
larger than the oscillation period on average \citep{Nechaeva19}.
Conversely, the decayless oscillation usually shows small but
weakly-decay displacement amplitude, which is less than the minor
radius of the oscillating loop
\citep{Wang12,Anfinogentov13,Duckenfield18,Mandal22}. The
oscillation periods of standing kink waves are measured from
sub-minute to dozens of minutes, and they are linearly increasing
with the loop lengths
\citep[e.g.,][]{Anfinogentov15,Nechaeva19,Zhang22,Petrova23,Zhong23}.

It is well accepted that the kink oscillation should be associated
with some external eruptions on the Sun \citep{Nakariakov21}. The
decaying kink oscillation is easily found to be driven by an
impulsive driver, for instance, an extreme-ultraviolet (EUV) wave, a
solar flare, a flux rope, and a coronal jet
\citep[e.g.,][]{Zimovets15,Shen18,Shen19a,Shen19b,Reeves20}.
However, the decayless kink oscillation appears to show no apparent
association with the solar transient
\citep[e.g.,][]{Gao22,Zhong22a,Li23}. On the other hand, the
decayless kink oscillation is nearly omnipresent in the solar
atmosphere \citep{Tian12,Li22a}, namely, they are frequently
observed in the coronal loop \citep[e.g.,][]{Li20,Safna22}, the
prominence thread \citep[e.g.,][]{Arregui18,Li18a}, the hot flare
loop \citep[e.g.,][]{Li18b,Shi23}, and the coronal bright point
\citep[e.g.,][]{Gao22}. Therefore, they could provide ongoing energy
to support the energy loss of the solar corona, which are believed
to be crucial for the coronal heating \citep{Van20,Yuan23}.
Obviously, the decayless kink oscillation should also have
continuous external drivers to supply that counteracts damping
\citep{Zhong22a,Nakariakov21}. In order to answer this issue,
several models or mechanisms have been proposed, such as the
self-oscillatory model \citep{Nakariakov16,Karampelas20},
random-motion excitation \citep{Ruderman21}, or p-modes exciter
\citep{Gao23}. A series of magnetohydrodynamic (MHD) simulations and
theoretical calculations have demonstrated that these models can
lead to decayless kink oscillations
\citep{Nakariakov16,Karampelas20,Ruderman21,Gao23}. However,
capturing observational evidence of continuously operating drivers
of decayless kink oscillations is still rare.

Quasi-periodic pulsations (QPPs) are frequently observed in the
light curves of solar flares and often associated with MHD waves in
solar atmospheres \citep[see][for a recent review]{Zimovets21}.
These phenomena could correspond to the quasi-periodic energy
release process \citep[e.g.,][]{Zimovets21b,Li22r}. A typical QPP is
generally characterized by a series of regular and repeated pulses
and it could be observed in multiple wavelengths, such as radio,
H$\alpha$, Ly$\alpha$, extreme-ultraviolet (EUV), soft and hard
X-rays (SXR/HXR), and even $\gamma$-rays. The quasi-periods are
measured from sub-seconds to several hundreds seconds
\citep[e.g.,][]{Nakariakov10,Kolotkov18,Kashapova20,Li21,Li22,Karlicky22,Kou22,Zhao23}.
The kink oscillation is often observed as the spatial displacement
disturbance of loop-like structures, while the QPP refers to the
periodic variation of light curves during a solar eruption.
Therefore, the kink-mode wave is commonly used to interpret the
observed QPP \citep[e.g.,][]{Nakariakov10,Li22b}. However, the loop
oscillation that is strongly associated with flare QPPs is rarely
reported. In this letter, we investigate the decayless kink
oscillation of a coronal loop, which could be excited by the
repetitive energy releases behind flare QPPs.

\section{Observation}
We analyzed a coronal loop that was associated with a solar flare
occurred on 2023 June 20, which lied in the active region of
NOAA~13234 near the solar east limb, namely, S17E73. They were
simultaneously measured by the Hard X-ray Imager
\citep[HXI;][]{Su19,Zhang19} on board the Advanced Space-based Solar
Observatory\footnote{http://aso-s.pmo.ac.cn/sodc/dataArchive.jsp}
\citep[ASO-S;][]{Gan23}, Konus-Wind \citep[KW;][]{Lysenko22},
Geostationary Operational Environmental Satellite (GOES), Expanded
Owens Valley Solar Array \citep[EOVSA;][]{Gary11}, STEREO/WAVES
(SWAVES), Atmospheric Imaging Assembly \citep[AIA;][]{Lemen12} on
board the Solar Dynamics Observatory (SDO), and the Chinese
H$\alpha$ Solar Explorer\footnote{https://ssdc.nju.edu.cn}
\citep[CHASE;][]{LiC22,Qiu22}. Figure~\ref{over} shows the light
curves in multiple wavelengths during the solar flare. The SXR
fluxes at 1-8~{\AA} (red) and 0.5-4~{\AA} (blue) are recorded by
GOES-18, which has a time cadence of 1~s.

ASO-S/HXI is designed to image solar flares in the HXR energy range
of about 15-300~keV. Its time cadence is 4~s in regular observation
mode and can be as high as $\sim$0.125~s in burst mode. In this
study, we use the full-disk light curve of three total flux monitors
(D92, D93, D94) in the range of 20-80~keV interpolated at a time
cadence of 1~s from the full cadence light curve, as shown by the
magenta line in Figure~\ref{over}~(b). KW is used for investigating
$\gamma$-ray bursts and solar flares, which works in two modes:
waiting and triggered modes. The count rate light curve has an
accumulation time of 2.944~s in the waiting mode, while it has a
varying time resolution (e.g., 2-256~ms) in the triggered mode.
Therefore, we interpolate the KW flux at 20-80~keV into an uniform
cadence of 2.944~s, as indicated by the green line in
Figure~\ref{over}~(b). We also use the radio dynamic spectra
measured by EOVSA and SWAVES, as shown by the background images in
Figure~\ref{over}. SWAVES acquires the radio spectrum with a time
cadence of 60~s, and it covers a frequency range of roughly
0.05-16.025~MHz. EOVSA is a microwave radioheliograph, which
provides the solar spectrum at frequencies of $\sim$1-18~GHz, and
the time cadence could be as high as 1~s. We note that some data
gaps appear in the EOVSA spectrum.

SDO/AIA takes full-disk solar maps in multiple EUV/UV passbands, and
the time cadence of seven EUV passbands is 12~s, while that of two
UV passbands is 24~s \citep{Lemen12}. In this study, we analyze AIA
maps in seven passbands of 131~{\AA} ($\sim$10~MK), 94~{\AA}
($\sim$6.3~MK), 193~{\AA} ($\sim$20~MK \& $\sim$1.6~MK), 211~{\AA}
($\sim$2.0~MK), 171~{\AA} ($\sim$0.63~MK), 1600~{\AA}
($\sim$0.1~MK), and 1700~{\AA} ($\sim$0.005~MK), as shown in
Figure~\ref{img}. All the AIA maps are pre-processed by
aia\_prep.pro, and have a same spatial resolution of 1.2\arcsec.
CHASE provides the spectroscopic observation of the full Sun in
wavebands of H$\alpha$ and \fei\ \citep{LiC22}. Here, we use the
spectral images at channels of H$\alpha$~6562.8~{\AA} and
\fei~6569.2~{\AA}, which mainly form in the solar chromosphere and
photosphere, respectively. Each spatial pixel corresponds to
$\sim$1.04\arcsec, and the time cadence is about 71~s.

\section{Results}
\begin{figure}
\centering
\includegraphics[width=0.9\linewidth,clip=]{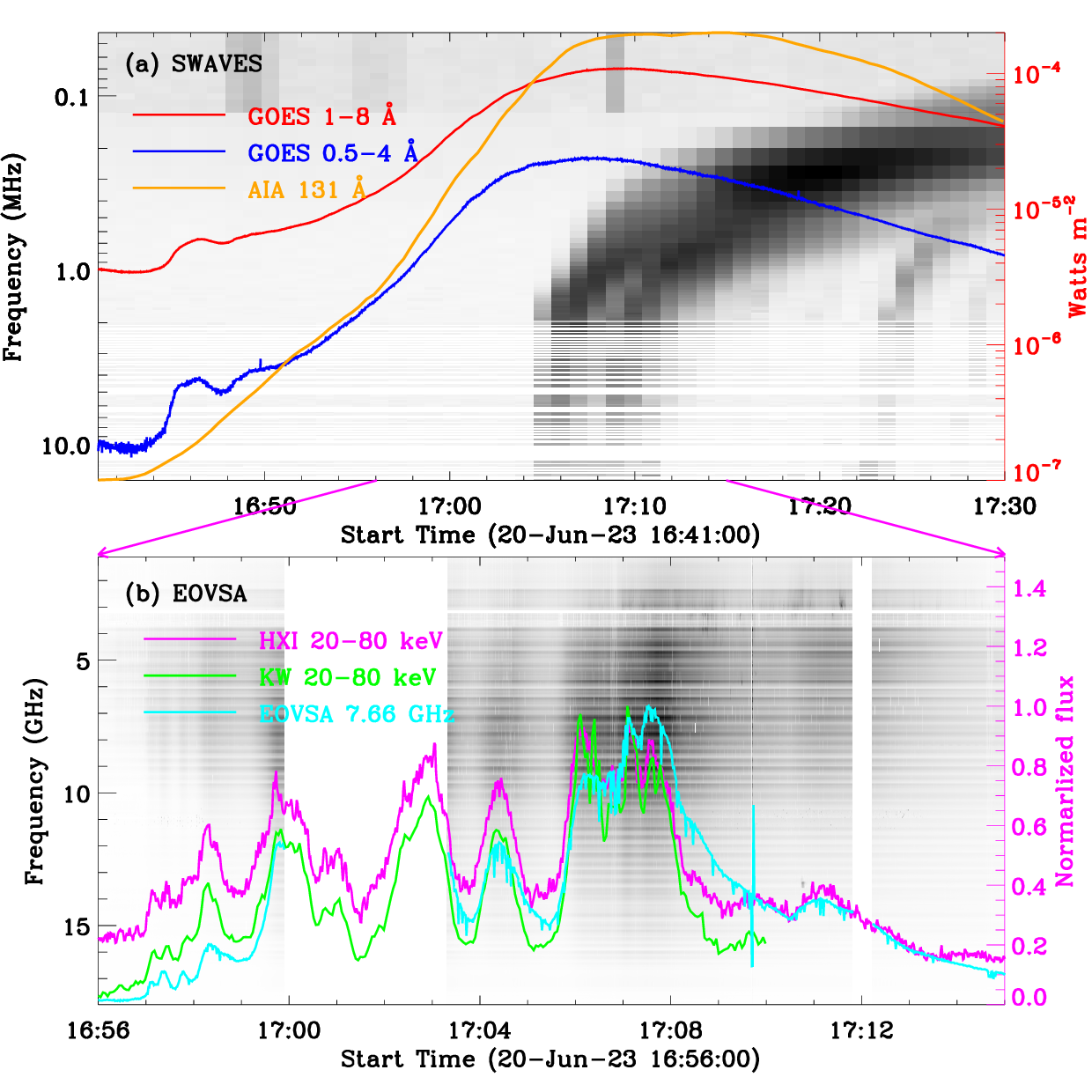}
\caption{Overview of the solar flare on 2023 June 20. a: Full-disk
light curves from 16:41~UT to 17:30~UT recorded by GOES at 1-8~{\AA}
(red) and 0.5-4~{\AA} (blue), and the local light curve integrated
over the flare region measured by SDO/AIA at 131~{\AA}. b: Full-disk
light curves between 16:56~UT and 17:15~UT in passbands of
HXI~20-80~keV (magenta), KW~20-80~keV (green), and EOVSA~7.66~GHz
(cyan). The context images are radio dynamic spectra observed by
SWAVES (a) and EOVSA (b), respectively. \label{over}}
\end{figure}

Figure~\ref{over}~presents the solar flare observed in passbands of
SXR, HXR and radio and microwave emissions. Panel~(a) shows
full-disk light curves at GOES~1-8~{\AA} (red) and 0.5-4.0~{\AA}
(blue) from 16:41~UT to 17:30~UT. The GOES flux indicates an
X1.1-class flare, which begins at $\sim$16:42~UT, and reach its
maximum at $\sim$17:09~UT. The orange line represents the local EUV
flux in the wavelength of AIA~131~{\AA}, which is integrated over
the flare region. Figure~\ref{over}~(b) shows HXR and microwave
fluxes during 16:56-17:15~UT measured by ASO-S/HXI (magenta), KW
(green), and EOVSA (cyan), respectively. Those light curves match
well with each other, and they all reveal at least five pulses,
which could be regarded as flare QPPs. The background images are
radio dynamic spectra observed by SWAVES (a) and EOVSA (b), which
show type III radio bursts at the lower and higher frequency ranges,
suggesting that nonthermal electrons are accelerated via magnetic
reconnections during the flare impulsive phase.

\begin{figure}
\centering
\includegraphics[width=0.9\linewidth,clip=]{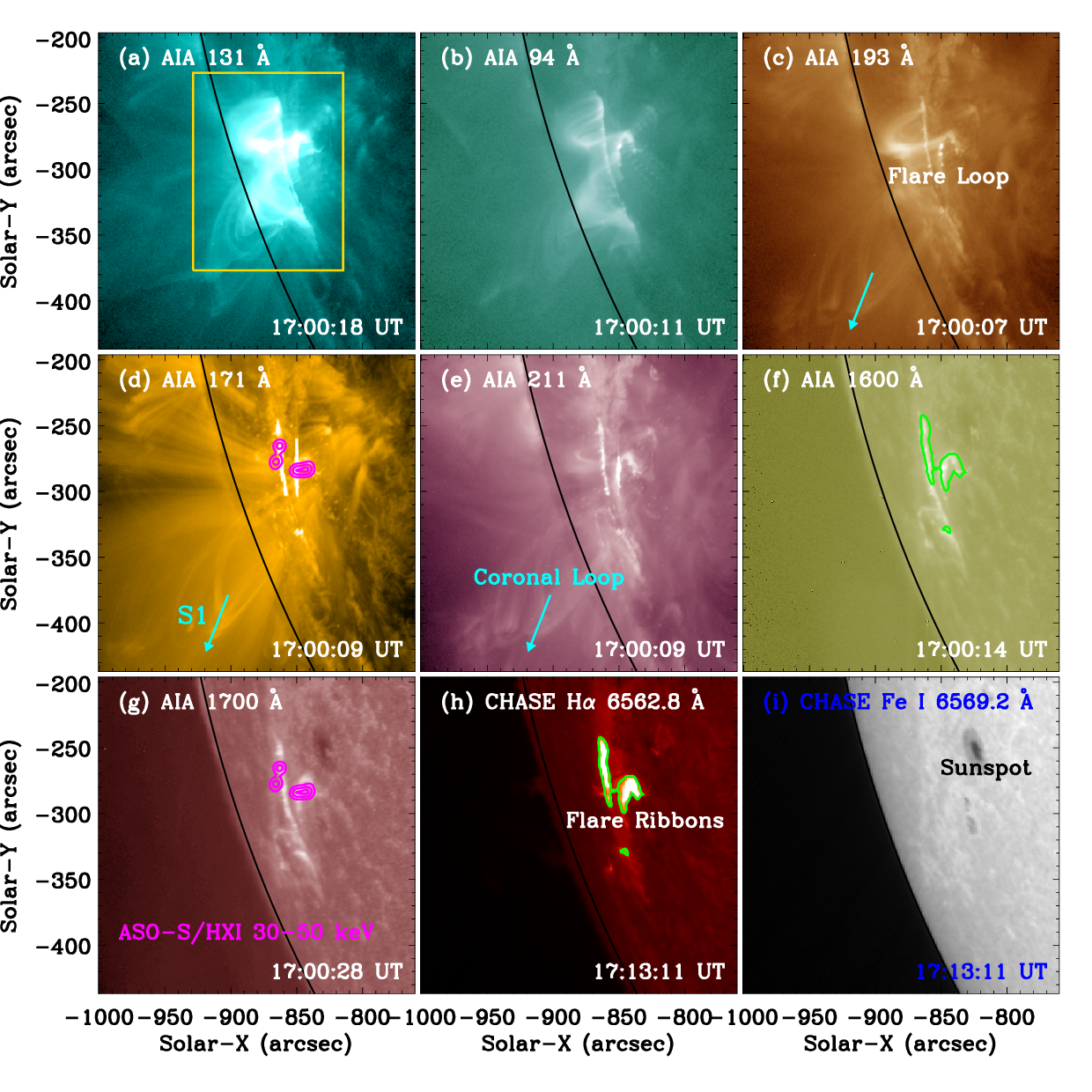}
\caption{Multi-wavelength snapshots with a FOV of
$\sim$240\arcsec$\times$240\arcsec\ measured by SDO/AIA at 131~{\AA}
(a), 94~{\AA} (b), 193~{\AA} (c), 171~{\AA} (d), 211~{\AA} (e),
1600~{\AA} (f), and 1700~{\AA} (g), and captured by CHASE in
passbands of H$\alpha$~6562.8~{\AA} (h) and \fei~6569.2~{\AA} (i),
respectively. The gold box outlines the flare region used to
integrate the local flare flux. The cyan arrow indicates the
targeted coronal loop, which is used to generate the time-distance
map. The magenta contours represents the HXR emission at
HXI~30-50~keV, and the contour levels are set 30\%, 60\%, and 90\%.
The green contours are derived from the H$\alpha$ radiation measured
by CHASE. An animation that shows the evolution of the solar flare
and coronal loop is available online.  \label{img}}
\end{figure}

Figure~\ref{img} shows the multi-wavelength images with a field of
view (FOV) of $\sim$240\arcsec$\times$240\arcsec. Panels~(a)-(c)
plot EUV maps in high-temperature wavelengths of AIA~131~{\AA},
94~{\AA}, and 193~{\AA}, which display some hot flare loops. The
gold rectangle outlines the flare region used to integrate the local
flux at AIA~131~{\AA} in Figure~\ref{over}~(a). While panels~d-h
illustrate the EUV/UV and H$\alpha$ maps in passbands of
AIA~171~{\AA}, 211~{\AA}, 1600~{\AA}, 1700~{\AA}, and
CHASE~6562.8~{\AA}, they all exhibit double flare ribbons and are
spatially correlated with two footpoints seen in the HXR emission,
as outlined by the green and magenta contours. The HXR map during
17:04:07-17:04:37~UT is reconstructed by the HXI\_CLEAN method
(pattern-based CLEAN algorithm for HXI), utilizing the detectors
from D19 to D91, namely, the subcollimator group G3 to G10, with a
spatial resolution of about 6.5\arcsec. We exclude the fine grids of
G1 and G2 since they are not calibrated yet and the fine structures
are not the focus of this study. It should be pointed out that ASO-S
was in calibration mode for other payloads during this flare and was
therefore pointing away from the solar disk center. The HXI pointing
data was regenerated using a machine-learning method. Panel~i
presents the CHASE map in the wavelength of \fei~6569.2~{\AA}, and
we can see a sunspot, but no signature of the flare radiation. It is
interesting that a bunch of coronal loops can be found in passbands
of AIA~171~{\AA}, 211~{\AA}, and 193~{\AA}, as indicated by the cyan
arrow.

\begin{figure}
\centering
\includegraphics[width=0.9\linewidth,clip=]{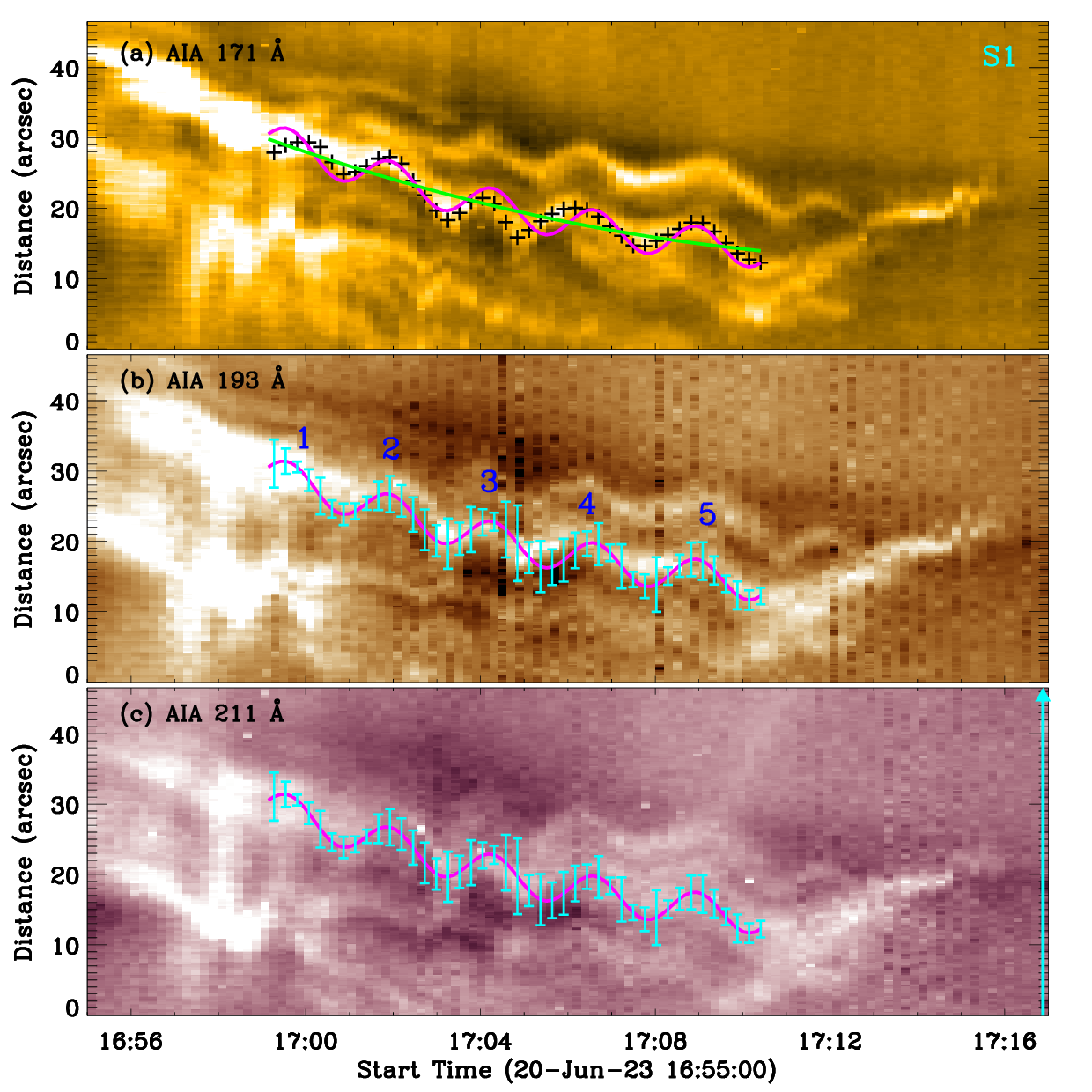}
\caption{Time-distance maps show the transverse oscillation of the
coronal loop at slit S1 in passbands of AIA~171~{\AA} (a), 193~{\AA}
(b), and 211~{\AA} (c). The pluses (`+') outline bright centers of
the oscillating loop. The magenta curve and the cyan error bars
represent the best-fitting result and their fitting uncertainties,
whereas the green line indicates the background trend. The Arabic
numerals mark five cycles of the loop oscillation. The cyan arrow
indicates the slit direction. \label{slit1}}
\end{figure}

The online animation (anim.mp4) shows the evolution of the solar
flare and the associated coronal loops. It can be seen that the
coronal loops appear transverse oscillations follow the flare
eruption. In order to capture the appearance of transverse
oscillations, one artificial straight slit (S1), which is nearly
perpendicular to the loop axis, is selected to generate the
time-distance (TD) maps. The cut slit is chosen at the position that
is close to the apparent loop apex, where is less overlapping with
the neighboring loops, as marked by the cyan arrow in
Figure~\ref{img}. Figure~\ref{slit1} presents the TD maps at slit S1
in passbands of AIA~171~{\AA}, 193~{\AA,} and 211~{\AA}. We can
immediately notice that several transverse oscillations appear in
these TD maps. Herein, there is only one transverse oscillation that
shows five apparent peaks analyzed, since it can be clearly seen in
three AIA passbands. Here, the peak of loop oscillation refers to
the maximum displacement of transverse oscillations in the TD map.
Similarly to previous studies \citep[e.g.,][]{Mandal21,Zhong22}, the
oscillation positions are determined from the bright centers of the
coronal loop at AIA~171~{\AA} by using the Gaussian fitting method,
as marked by the black pluses (`+') in panel~(a). The transverse
oscillation does not show any apparent decay, but it drifts
nonlinearly in the plane-of-sky. Therefore, the combination of a
sine function and a nonlinear trend is applied to fit the loop
oscillation \citep[e.g.,][]{Anfinogentov15,Li20,Gao22,Li23}, as
shown by Equation~\ref{eq1}:

\begin{equation}
  A(t)=A_m \cdot \sin(\frac{2 \pi}{P}~t+ \psi ) + f(t),
\label{eq1}
\end{equation}
\noindent Here, $A_m$ represents the displacement amplitude, $P$ is
the oscillation period, and $\psi$ refers to the initial phase,
while $f(t)$ stands for the second-order polynomial approximation.
Next, the velocity amplitude ($v_m$) is obtained by the derivative
of the displacement amplitude \citep[cf.][]{Li22a,Li23a,Petrova23}.
The overplotted magenta curve in Figure~\ref{slit1} is the
best-fitting result with Equation~\ref{eq1} and it appears to agree
with the oscillation positions (`+'). The green line in panel~a
represents a nonlinear trend of the oscillating loop, which is
derived from a second-order polynomial approximation. We want to
state that the magenta curve in panels~b and c is exactly the same
as that in panel~a and it seems to be not a good fit in passbands of
AIA~193~{\AA} and 211~{\AA}. On the other hand, they appear to match
with each other when we consider the fitting uncertainty, as shown
by the cyan error bars.

\begin{figure}
\centering
\includegraphics[width=0.9\linewidth,clip=]{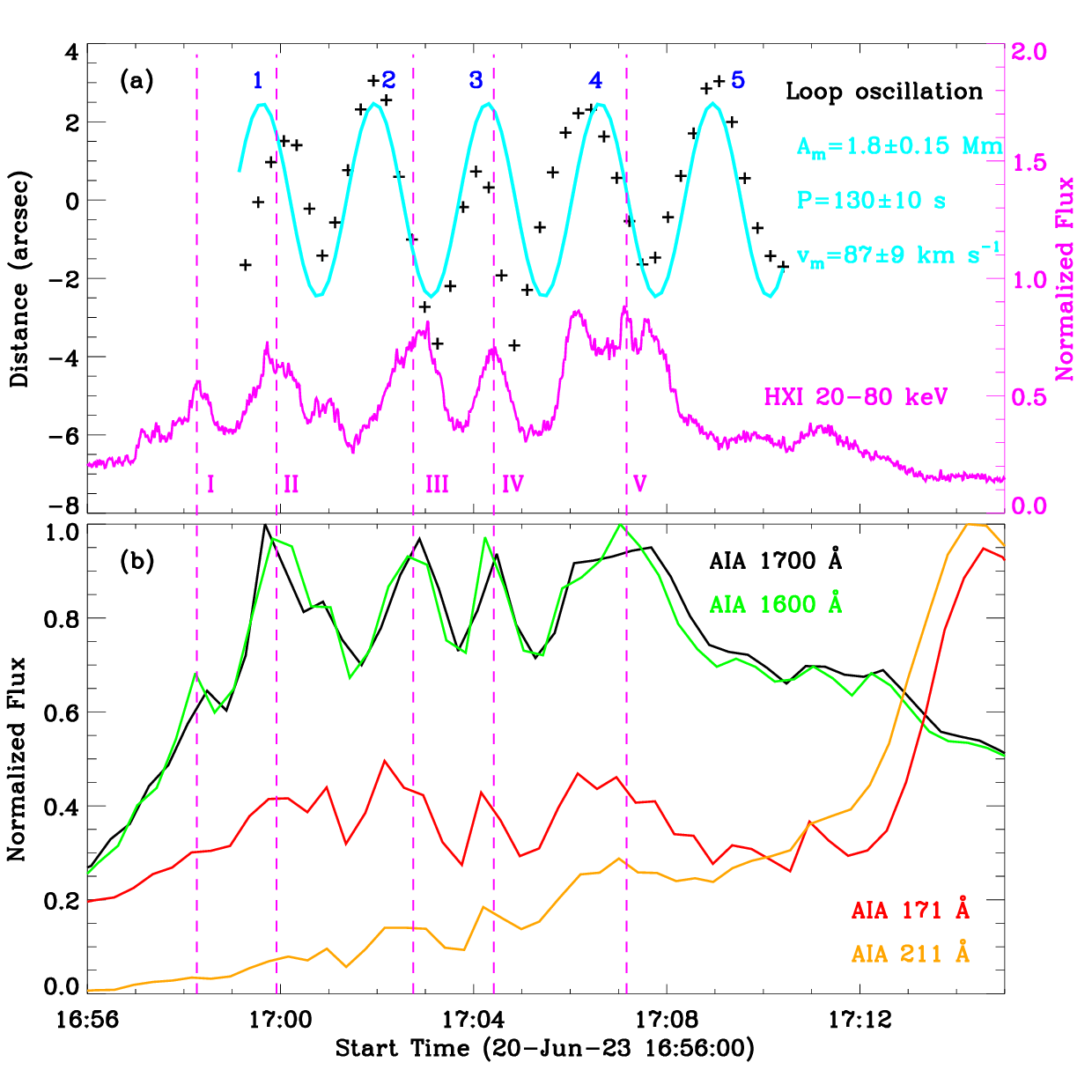}
\caption{Oscillating positions (`+') after removing the background
trend and its best-fitting result, shown in cyan (a). They are taken
from the coronal loop at AIA~171~{\AA}, as shown in
Figure~\ref{slit1}. The magenta curve shows the HXR flux measured by
ASO-S/HXI at 20-80~keV during 16:56-17:15~UT. Local light curve
integrated over the flare region (indicated by the gold box in
Figure~\ref{img}) in psaabands of AIA~1700~{\AA} (black),
1600~{\AA}(green), 171~{\AA} (red), and 211~{\AA} (orange),
respectively (b). The Arabic and English numbers, as well as the
dashed vertical lines outline these oscillating peaks. \label{flux}}
\end{figure}

Figure~\ref{flux}~(a) presents the oscillating positions (`+') of
the coronal loop and the best-fitting result (cyan) after removing
the nonlinear trend ($f(t)$), the fitting parameters and their
uncertainties such as the oscillation period, displacement and
velocity amplitudes are also labeled. The normalized HXR flux at
HXI~20-80~keV is also drawn, as shown by the magenta curve. One can
immediately notice that both the loop oscillation and the HXR flux
reveal at least 5 cycles, as marked by the Arabic and English
numbers. The cycle of each HXR pulse appears earlier than that of
the loop oscillation, and a time difference of about 110~s was
estimated via cross correlation. Those observational facts imply
that the transverse oscillation of the coronal loop could be
strongly associated with the flare QPP in the HXR channel. We also
notice that the HXR flux appears much more small sub-peaks, for
instance, the HXR pulse "V" contains three sub-peaks, which might be
due to the high time resolution of ASO-S/HXI. We then plot local
light curves integrated over the flare region (gold rectangle in
Figure~\ref{img}), as shown in Figure~\ref{flux}~(b). Obviously, the
light curves in passbands of AIA~1700~{\AA}~(black) and 1600~{\AA}
(green) display five apparent peaks, which are almost synchronous
with the HXR pulses, as indicated by the dashed vertical lines. On
the other hand, the light curves at AIA~171~{\AA}~(black) and
211~{\AA} (green) also show five main peaks, while some main peaks
also contain sub-peaks, similarly to what has observed in the HXR
flux; this suggests the coexistence of multiple periodicities in the
flare emission.

\begin{figure}
\centering
\includegraphics[width=0.9\linewidth,clip=]{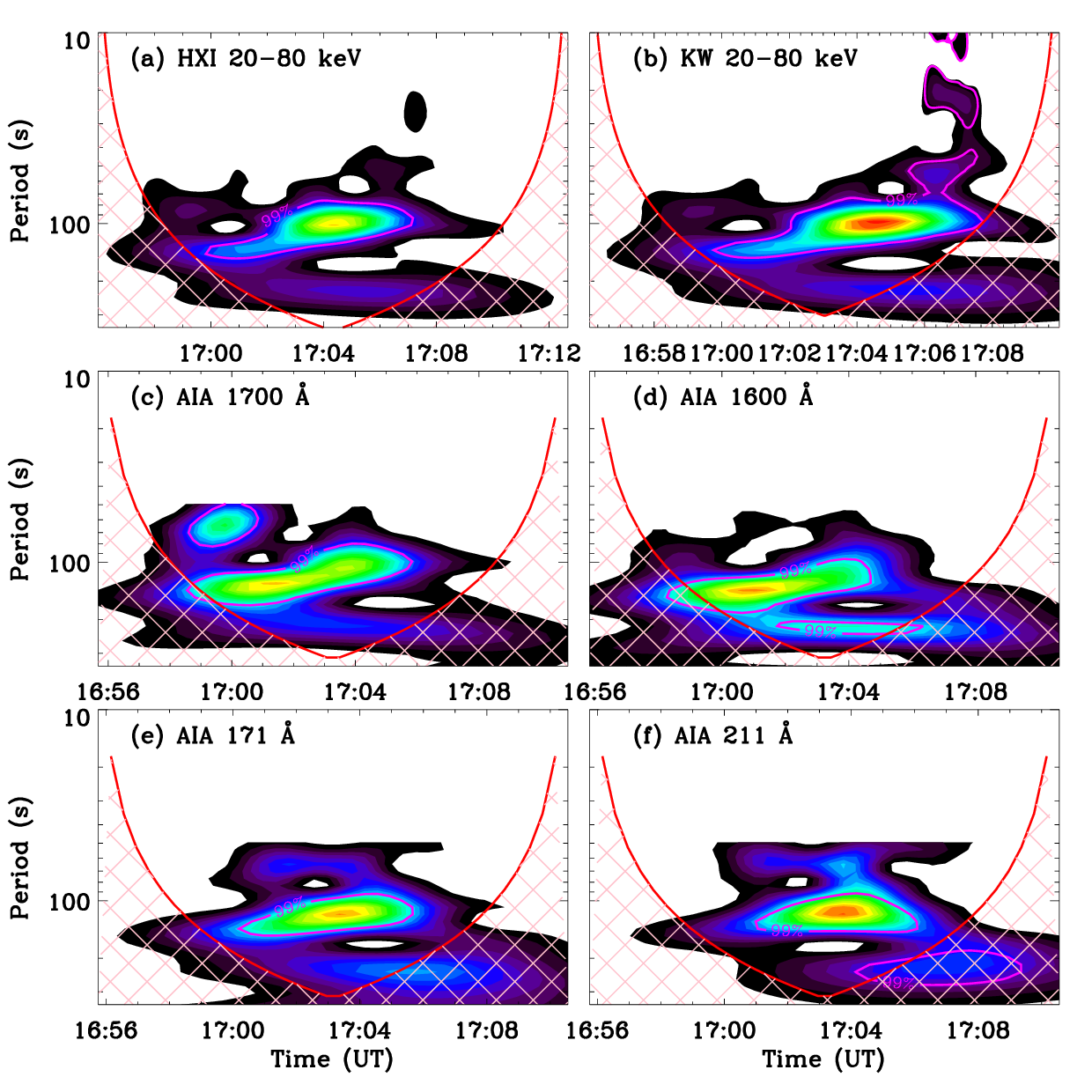}
\caption{Morlet wavelet power spectra, as seen in passands of
HXI~20-80~keV (a), KW~20-80~keV (b), AIA~1700~{\AA} (c) and
1600~{\AA} (d), 171~{\AA} (e), and 211~{\AA} (f). The magenta
contours indicate a significance level of 99\%. Anything outside the
red curve is dubious. \label{wav}}
\end{figure}

In order to identify the quasi-period of flare QPPs, the wavelet
transform with the "Morlet" mother function \citep{Torrence98} was
applied for the detrended light curves after removing a $\sim$180~s
running average \citep{Tian12,Li18b}, since we want to enhance the
short-period oscillation and suppress the long-period trend.
Figure~\ref{wav} shows the Morlet wavelet power spectra in passbands
of HXI~20-80~keV (a), KW~20-80~keV (b), AIA~1700~{\AA} (c),
1600~{\AA} (d), 171~{\AA} (e), and 211~{\AA} (f). They all show an
enhanced power over a broad range of quasi-periods, namely, a
periodicity range of about 100-130~s. We want to state that the
quasi-periods refer to the enhanced power range inside the 99\%
significance level. The quasi-periods agree with the average period
(i.e., $\sim$130$\pm$10~s) of the loop oscillation, confirming that
the transverse oscillation of the coronal loop could be strongly
associated the flare QPPs. We did not perform the wavelet transform
for the radio data measured by EOVSA because it has some data gaps,
resulting into a discontinuous microwave flux.

\section{Discussion}
Transverse oscillations are commonly detected in loop-like
structures, such as coronal and flare loops, prominence or filament
threads, and even umbral fibrils
\citep[e.g.,][]{Anfinogentov15,Nechaeva19,Li20,Li22a,Zhang22,Yuan23}.
In this letter, we study the transverse oscillation of a coronal
loop, which is perpendicular to the loop axis at the
apparent loop apex. It shows five significant cycles and reveals no
apparent decay in the displacement amplitude, which could be
regarded as the decayless kink oscillation
\citep{Tian12,Anfinogentov15,Li23}. The observed kink oscillation
shows an average period of $\sim$130$\pm$10~s, which is consistent
with previous observations that were in the range of tens to
hundreds of seconds
\citep{Anfinogentov15,Li20,Mandal21,Shi22,Zhong23}. While it has a
large displacement amplitude, which is measured to be about
1.8$\pm$0.15~Mm. Such a large displacement amplitude is rarely
reported in decayless oscillations. Previous observations found that
their displacement amplitude have mostly been less than 0.5~Mm
\citep{Anfinogentov15,Gao22} and the maximal displacement amplitude
of 1.16~Mm was detected in decayless kink oscillations induced by a
solar flare \citep[cf.][]{Mandal21}. \cite{Li20} also reported a
decayless kink oscillation with an amplitude of 0.8~Mm. On the other
hand, our velocity amplitude is estimated to be 87~km~s$^{-1}$ and
it is much larger than previous statistical studies for decayless
oscillations within longer periods, namely, $\sim$1-8~km~s$^{-1}$
\citep{Tian12,Gao22}. Our velocity amplitude is similar to those of
decayless oscillations with shorter periods
\cite[e.g.,][]{Petrova23,Li23}.

A solar flare occurs near the footpoint of the oscillating loop. It
shows five repetitive and successive pulses at HXR 20-80~keV
measured by ASO-S/HXI and KW, which could be regarded as the
signature of flare QPPs. It has a broad range of quasi-periods, such
as about 100-130~s. The flare QPPs within similar quasi-periods are
also observed in EUV/UV passbands of AIA~1700~{\AA}, 1600~{\AA},
171~{\AA}, and 211~{\AA}, and these AIA channels show double flare
ribbons, which superposed on two HXR footpoints. While those AIA
channels at AIA 131~{\AA}, 94~{\AA} and 193~{\AA} do not show the
similar QPP feature, since they reveal flare loops. Moreover, some
HXR/UV pulses appear to match well with the radio peaks observed by
EOVSA, implying the presence of QPP in the microwave emission. Our
observations suggest that the flare QPPs are probably caused by
repetitive magnetic reconnections
\citep[e.g.,][]{Kupriyanova20,Zimovets21}, which could periodically
accelerate nonthermal electrons that precipitate into flare
footpoints or ribbons. The repeated magnetic reconnections will lead
to quasi-periodic energy release processes, namely, five energy
releases during the X1.1 flare.

It is interesting that the same periodicity is found between the
decayless oscillation and flare QPPs. Moreover, each cycle of the
flare QPPs appears earlier than that of the decayless oscillation.
Thus, the repetitive energy releases of the X1.1 flare could be
considered as continuously operating drivers of the decayless kink
oscillation. The large displacement amplitude could be attributed to
the X1.1 flare, which could release a large amount of energies. This
is different from previous observations of large-amplitude decayless
oscillations that were triggered by small flares \citep{Mandal21} or
small reconnection events \citep{Li20}. \cite{Mandal21} found that
the solar flare could only increase oscillation amplitudes but did
not change the oscillation nature \citep[see also][]{Shi22}.
However, these studies did not show a distinct one-to-one
correspondence over time. In our case, the energy flow appears as
"an external driver" induced by a single flare energy release,
leading to a large-amplitude kink oscillation of the coronal loop,
followed by plasma heating caused by rapid damping (i.e., one
oscillation cycle). This process is repeated five times, since the
solar flare shows five energy releases via repetitive magnetic
reconnections. This model could also explain the small-amplitude
decayless oscillation, supposing that the external drivers are
continuously existing. However, those external drivers are difficult
to observe, mainly because of their fine scales.

Here, we assume the similar damping mechanism in decayless
oscillations with the decaying oscillations and it is compensated
with continuous energy supply from the ongoing flare energy
releases. Then, we can construct an equation like:
\begin{eqnarray}
  E(t) = \frac{1}{2}(\rho_i~v_m^{2}+\frac{b^2}{\mu_0}) \cdot e^{-t/\tau},
\end{eqnarray}
where $E(t)$ represents the energy (kinetic + magnetic) density
averaged over the oscillation cycle, $\tau,$ is the decaying time
measured for the displacement amplitude, while $v_m$, $\rho_i$, and
$b$ are the velocity amplitude, plasma density and magnetic field
perturbation, respectively. Then, its derivative $\frac{dE(t)}{dt}$
at t=0 will give us an estimation for the oscillation energy losses
($\varepsilon$):
\begin{eqnarray}
  \varepsilon = \frac{dE(t)}{dt} = \frac{1}{2\tau} (\rho_i~v_m^{2}+\frac{b^2}{\mu_0}),
\end{eqnarray}

In this study, we know that the decayless oscillation has five
oscillation cycles, which correspond to five-cycle energy releases
from the flare. If the oscillating loop is affected by a periodic
force with the oscillation period, the effect of the resonance must
take place and the oscillation amplitude would grow over time
\citep{Nakariakov09}. However, the observed amplitude remains
constant, suggesting that the wave energy of each oscillation cycle
could completely dissipate at a time scale shorter than or
approximately equal to one oscillation period, namely,
$\tau~\leq~P$. According to the resonant absorption theory
\citep{Goossens02}, such a short decay time indicates an abnormal
ratio of the loop radius and thickness of the transition layer or an
abnormal density ratio inside and outside the loop, either too dense
or too rarified. Possible supporting observational evidence to this
interpretation is the half-cycle transverse perturbation of the
outer loop in \cite{Ning22}. In this work, measuring the loop
density may answer this concern but out of the scope of this work.
Nevertheless, additional observations and numerical simulations in
this direction are necessary for further investigation in the
future.

\section{Summary}
Using the observations from SDO/AIA, ASO-S/HXI, KW, CHASE, EOVSA,
SWAVES, and GOES, we investigated an apparent decayless kink
oscillation and the associated flare QPPs. Our main conclusions are
summarized as follows:

(1) The apparent decayless kink oscillation can be simultaneously
seen in passbands of AIA~171~{\AA}, 193~{\AA}, and 211~{\AA},
indicating that the oscillating loop is multi-thermal in nature. The
oscillation period is measured to be about 130$\pm$10~s and the
displacement amplitude is as large as $\sim$1.8$\pm$0.15~Mm.

(2) The flare QPPs are simultaneously detected in passbands
of HXI~20-80~keV, KW~20-80~keV, AIA~1700~{\AA}, 1600~{\AA},
171~{\AA}, and 211~{\AA} during the flare impulsive phase, which
might be caused by repetitive magnetic reconnections. The
quasi-period is estimated to be $\sim$100-130~s. The large period
range implies the co-existence of multiple periodicities.

(3) The apparent decayless kink oscillation and flare QPPs share the
same quasi-period and a significant time difference of about 110~s
can be seen between them. Those observational facts provide
sufficient evidences that the decayless kink oscillation could be
driven by continuously operating energies released from the X1.1
flare.

(4) We propose that the repetitive energy releases behind the flare
QPPs could trigger the kink oscillation intermittently for five
times, making it apparent decayless. In this interpretation, each
cycle of the kink oscillation is assumed to decay rapidly in less
than one oscillation period.

\begin{acknowledgements}
The authors would like to thank the referee for valuable comments
and suggestions. This work is funded by the National Key R\&D
Program of China 2022YFF0503002 (2022YFF0503000), NSFC under grants
11973092, 12073081, 12333009, 12333010, 11820101002. D. Li is
supported by the Surface Project of Jiangsu Province (BK20211402)
and the Specialized Research Fund for State Key Laboratories. Z. T.
Li is supported by the Prominent Postdoctoral Project of Jiangsu
Province. This work is also supported by the Strategic Priority
Research Program of the Chinese Academy of Sciences, Grant No.
XDB0560000. ASO-S mission is supported by the Strategic Priority
Research Program on Space Science, the Chinese Academy of Sciences,
Grant No. XDA15320000. The CHASE mission is supported by China
National Space Administration (CNSA).
\end{acknowledgements}

\end{document}